\documentclass[11pt]{article}
\usepackage{graphicx}
\usepackage[table]{xcolor}% http://ctan.org/pkg/xcolor
\usepackage{amsmath}
\usepackage{array,booktabs,siunitx}
\usepackage[utf8]{inputenc}
\usepackage{multirow}
\renewcommand{\arraystretch}{1.8}
\usepackage{fmtcount} % for \ordinalnum
\usepackage{array,multirow}
\newcolumntype{C}[1]{>{\centering\arraybackslash}m{#1}}
\usepackage{caption}
\usepackage{hyperref}
\usepackage{fancyhdr}

\usepackage{titling}

\usepackage{amsmath,amssymb}
\usepackage{slashed}
\usepackage{xcolor} % Define colors
\usepackage{graphicx}
\usepackage{dcolumn} % Align table columns on decimal point
\usepackage{bm} % Bold math
\usepackage{epsfig}
\usepackage{epstopdf}
\usepackage{grffile}
\usepackage{color}
\usepackage{colordvi}
\usepackage{amsmath,amssymb}
\usepackage{rotating}
\usepackage{lscape}
\usepackage{cite}
\usepackage{float}
\usepackage{hyperref}
\usepackage{url}
\usepackage{graphicx}
\usepackage{color}
\usepackage{amssymb}
\usepackage{amsmath}
\usepackage{mathtools}
\usepackage{bm}	
\usepackage{flexisym}
\usepackage{amssymb}
\usepackage{physics}
\usepackage{mathrsfs}
\usepackage{simplewick}
\usepackage{tikz}
\usetikzlibrary{arrows,shapes}
\usetikzlibrary{trees}
\usetikzlibrary{matrix,arrows} 			
\usetikzlibrary{positioning}			
\usetikzlibrary{calc,through}			
\usetikzlibrary{decorations.pathreplacing}  
\usepackage{pgffor}						
\usetikzlibrary{decorations.pathmorphing}
\usetikzlibrary{decorations.markings}
\tikzset{
	vector/.style={decorate, decoration={snake}, draw},
	provector/.style={decorate, decoration={snake,amplitude=2.5pt}, draw},
	antivector/.style={decorate, decoration={snake,amplitude=-2.5pt}, draw},
	fermion/.style={draw=black, postaction={decorate},
		decoration={markings,mark=at position .55 with {\arrow[draw=black]{>}}}},
	fermionbar/.style={draw=black, postaction={decorate},
		decoration={markings,mark=at position .55 with {\arrow[draw=black]{<}}}},
	fermionnoarrow/.style={draw=black},
	gluon/.style={decorate, draw=black,
		decoration={coil,amplitude=4pt, segment length=5pt}},
	scalar/.style={dashed,draw=black, postaction={decorate},
		decoration={markings,mark=at position .55 with {\arrow[draw=black]{>}}}},
	scalarbar/.style={dashed,draw=black, postaction={decorate},
		decoration={markings,mark=at position .55 with {\arrow[draw=black]{<}}}},
	scalarnoarrow/.style={dashed,draw=black},
	electron/.style={draw=black, postaction={decorate},
		decoration={markings,mark=at position .55 with {\arrow[draw=black]{>}}}},
	bigvector/.style={decorate, decoration={snake,amplitude=4pt}, draw},
}
\tikzstyle{block} = [draw, rectangle, minimum height=3em, minimum width=6em]
\usepackage{hyperref}

\usepackage{titling}
\usepackage{subcaption}
\newcommand{\subtitle}[1]{%
	\posttitle{%
		\par\end{center}
	\begin{center}\large#1\end{center}
	\vskip0.5em}%
}

\def\Re{{\cal R \mskip-4mu \lower.1ex \hbox{\it e}\,}}
\def\Im{{\cal I \mskip-5mu \lower.1ex \hbox{\it m}\,}}

\def\tev{\,{\ifmmode\mathrm {TeV}\else TeV\fi}}
\def\gev{\,{\ifmmode\mathrm {GeV}\else GeV\fi}}
\def\mev{\,{\ifmmode\mathrm {MeV}\else MeV\fi}}
\def\to{\rightarrow}

\begin{document}

\begin{center}

\vspace*{15mm}
\vspace{1cm}
{\Large \bf Exploring  Axion-Like Particle Couplings through Single Top $tW$-Channel and  Top Pair Production at the LHC}

\vspace{1cm}

{\bf Yasaman Hosseini$^{1}$ and Mojtaba Mohammadi Najafabadi$^{1,2}$ }\\
{\small\sl 
$^{1}$ School of Particles and Accelerators, Institute for Research in Fundamental Sciences (IPM) P.O. Box 19395-5531, Tehran, Iran\\
$^{2}$Experimental Physics Department, CERN, 1211 Geneva 23, Switzerland  \\
 }\vspace*{.2cm}
\end{center}

\vspace*{.2cm}

\vspace*{10mm}

%%%%%%%%%%%%%%%%%%%%%%%%%%%%%%%%  
\begin{abstract}\label{abstract}
In this paper, we explore the interactions between light axion-like particles (ALPs) and top quarks, gluons, 
and W-bosons. Utilizing high-energy probes at the LHC, we probe these couplings in the context of 
single top quark production associated with a W-boson and an ALP. We analyze the latest LHC Run II data 
on $t\bar{t}$ spin correlation measurements to establish constraints on these couplings. 
Additionally, we compare these constraints with those derived from loop-induced couplings 
and indirect searches involving B-meson decays. 
Furthermore, we derive two-dimensional limits on the ALP couplings with top quarks and gluons.
\end{abstract}
%%%%%%%%%%%%%%%%%%%%%%%%%%%%%%%%  

%\clearpage
\newpage

%%%%%%%%%%%%%%%%%%%%%%%%%%%%%%%%  
\section{Introduction}

Despite the remarkable achievements of the Standard Model (SM) of particle physics, 
certain observational and theoretical aspects remain unexplained within its framework. 
Examples include the existence of Dark Matter (DM), neutrino mass, and the baryon asymmetry. 
Theoretical challenges such as the strong CP and hierarchy problems also persist.
To address these shortcomings, numerous theories beyond the SM (BSM) 
have been proposed. Despite extensive efforts to detect signatures of these BSM theories 
at the Large Hadron Collider (LHC), no significant evidence of new physics at high energies 
has been found thus far. Consequently, there is growing interest in exploring the possibility 
of new light degrees of freedom or weakly coupled states that could augment the existing 
content of the SM.

Axion-like particles (ALPs) \cite{i1,i2} frequently arise in 
extensions of the SM.  They are often proposed as solutions to the strong CP problem 
or as manifestations of newly broken global symmetries in nature \cite{i3, i4, i5, i6, i7, i8, g1, g2}.
ALPs are under scrutiny in various experimental endeavors, 
including high-energy colliders  \cite{i9, i10, i11, i12, i13, i14, i15, i16, i17, i18, i19, i20, i21, i22, i23, i24, i25, i26, i27, i279, i2799, i950,i951, i952, g3}, 
 investigations into flavor physics \cite{i289, i29, i30, i31, i32, i33}, beam dump experiments \cite{i35, i36}, 
 and assessments of their astrophysical implications \cite{i37}.

The main theoretical framework utilized in ALP studies is the model-independent 
approach of effective field theories (EFT). In this methodology, the ALP decay constant $f_{a}$ serves as the scale of consideration for new physics. 
It regulates the impact of higher-dimensional operators constructed from both the SM fields and the ALP field $a$.
The comprehensive one-loop renormalization group evolution of all parameters within the 
EFT framework for axion-like particles up to dimension five, 
encompassing scales both above and below the electroweak scale, has been detailed in Ref. \cite{i38}.
In Ref. \cite{i39}, a global fit of the effective Lagrangian for ALPs to data is conducted. 
This fit combines LHC observables from top physics, dijet and diboson production with 
electroweak precision observables. Through this analysis, the full parameter space of ALPs has been thoroughly explored.
 
 The hierarchical nature of flavor in the couplings between the ALPs and fermions assigns a 
 unique significance to the heaviest observed fermion {\it i.e.} the top quark in ALP phenomenology.
Investigating the coupling between the ALP and top quarks at high energies provides insights into 
potential effects at lower energy scales. Notably, 
in new models where ALP couplings with gauge bosons arise from fermion loops, 
probing the ALP-top coupling ($a-t-\bar{t}$) implies a simultaneous exploration of the ALP couplings with gauge bosons.

In Ref. \cite{i40}, the ALP's interactions with both top quarks and gauge bosons are examined simultaneously by 
investigating its contributions to the top quark's magnetic and chromomagnetic dipole moments. 
This study demonstrates that probing the magnetic and chromomagnetic dipole moments of 
the top quark provides a unique avenue to explore the ALP's connections to both top quarks and gauge bosons concurrently.
Furthermore, the direct exploration of the ALP-top couplings via single top t-channel and top quark pair production at the LHC
has been discussed in Ref.\cite{i40} but no limits were presented.
In Ref. \cite{i41}, the interplay between the  $t\bar{t}+a$ signature and the off-shell production 
of the ALP leading to a $ t\bar{t}$ final state is investigated. It has been discovered that the $t\bar{t}+\text{MET}$
 signatures, where ALPs evade detection, exhibit heightened sensitivity to the ALP-top coupling 
 compared to the off-shell production of $t\bar{t}$ via ALPs. Furthermore, limits on the ALP-top coupling, 
 derived from the reinterpretation of SUSY searches for {\it stops} under the assumption of ALP collider-stability, 
 are provided in Ref.\cite{i41}.
In Ref. \cite{i42}, measurements of $t\bar{t}$ production, 
along with Higgs and meson decays, have been leveraged to probe the sensitivity of 
ALPs with respect to top quarks. Additionally, the study demonstrated that four-top production provides 
supplementary sensitivity to ALPs featuring top couplings over a broad mass spectrum.
In Refs.\cite{i17,i43}, future prospects for exploring the ALP-top coupling have been deliberated.

In this study, we investigate collider-stable ALPs that evade detection 
through associated production with a top quark and a W boson at the LHC. We utilize the 
differential distribution measured by the CMS experiment \cite{cmstw} at a center-of-mass energy of 13 TeV, 
with an integrated luminosity of 138 fb$^{-1}$, to establish constraints on the couplings of the ALP 
with top quarks, W bosons, and gluons.
Furthermore, we assess the sensitivity of spin correlation measurements in $t\bar{t}$
events to the ALP model parameter space. Constraints on the ALP couplings with top quarks 
and gluons are derived from CMS experiment measurements of spin correlation at a center-of-mass 
energy of 13 TeV, with an integrated luminosity of 35.6 fb$^{-1}$ \cite{cmsspin}.
Throughout this investigation, our focus centers on ALPs with a mass of 1 MeV, 
considering those that do not decay inside the detector and manifest as missing transverse momentum.

This paper is organized as follows. In Section \ref{sec:th}, we give a quick introduction of the theory behind ALPs. 
Section \ref{sectw} describes the details of search for ALPs using the LHC through $t+W+a$ production.
In Section \ref{secspin}, we dig into the sensitivity of the spin correlation in $t\bar{t}$ events to probe ALPs' interactions. 
 Bound of the ALP couplings from these studies are presented at confidence level of $95\%$. 
 Section \ref{comp} provides a comparison of constraints on ALP couplings from this study and other studies.
The two-dimensional limits on the ALP couplings with top quarks and gluons are discussed and presented in Section \ref{2Dlimit}.
 Finally, in Section \ref{secsummary}, we wrap up the paper with a summary and conclusions.

%%%%%%%%%%%%%%%%%%%%%%%%%%%%%%%%  
\section{Theoretical framework}
\label{sec:th}

Axion like particles are spin-0 particles that emerge in beyond standard models, incorporating an additional U(1) symmetry which is spontaneously broken. These particles are singlet under the standard model gauge group and odd under CP transformation. The interaction between ALPs, denoted as $a$, and the SM fields is described by an Effective Lagrangian which at the lowest order, it is expressed as\cite{i12}:
\begin{eqnarray}
	\mathcal{L}_{eff}^{D\leq5} &=& \mathcal{L}_{SM} + \dfrac{1}{2} (\partial^\mu a)(\partial_\mu a) - \dfrac{1}{2} m_a^2 a^2 \nonumber\\
	&-& c_{\tilde{W}} \dfrac{a}{f_a} W^a_{\mu\nu} \tilde{W}^{a\mu\nu} - c_{\tilde{B}} \dfrac{a}{f_a} B_{\mu\nu} \tilde{B}^{\mu\nu} - c_{\tilde{G}} \dfrac{a}{f_a} G^a_{\mu\nu} \tilde{G}^{a\mu\nu} \nonumber\\
	&+& c_{a\Phi} \dfrac{\partial^\mu a}{f_a} \left(\Phi^\dagger i \overleftrightarrow{D}_\mu \Phi\right) + \dfrac{\partial^\mu a}{f_a} \sum_{F} \bar{\Psi}_F \textbf{C}_F \Psi_F,
	\label{LagALP1} 
\end{eqnarray}
where $m_a$ is the mass of the ALP. The $\Phi$ represents the Higgs boson doublet and the $G^{\mu\nu}$, $W^{\mu\nu}$, $B^{\mu\nu}$ are field strengths for $SU(3)_c$$\times$$SU(2)_L$$\times$$U(1)_Y$, also the dual field strengths are defined as $\tilde{X}^{\mu\nu} \equiv \dfrac{1}{2} \epsilon^{\mu\nu\alpha\beta}X_{\alpha\beta}$, 
where $\epsilon^{\mu\nu\alpha\beta}$ is Levi-Civita symbol. The sum in the last line is performed over all the SM fermion fields $F = L_L, Q_L, e_R, d_R, u_R$, where $L_L$ and 
$Q_L$ are $SU(2)_L$ doublets, while $e_R, d_R, u_R$ are $SU(2)_L$ singlets. The $\textbf{C}_F$ are Hermitian matrices in flavor space. In this study, we exclusively focus 
on the bosonic Lagrangian and for simplicity  $\textbf{C}_F$ is assumed to be zero.
Applying the Higgs field redefinition:
\begin{equation}
	\Phi \rightarrow e^{i c_{a\Phi} a/f_a} \Phi,
\end{equation}
the Lagrangian in Eq.\ref{LagALP1} can be written as:
\begin{eqnarray}
	\mathcal{L}_{eff}^{D\leq5} &=& \mathcal{L}_{SM} + \dfrac{1}{2} (\partial^\mu a)(\partial_\mu a) - \dfrac{1}{2} m_a^2 a^2 \nonumber\\
	&-& c_{\tilde{W}} \dfrac{a}{f_a} W^a_{\mu\nu} \tilde{W}^{a\mu\nu} - c_{\tilde{B}} \dfrac{a}{f_a} B_{\mu\nu} \tilde{B}^{\mu\nu}  \tilde{G}^{a\mu\nu} \nonumber\\
	&-& c_{\tilde{G}} \dfrac{a}{f_a} G^a_{\mu\nu} + c_{a\Phi} \textbf{O}_{a\Phi}^{\psi}, 
	\label{LagALP2}
\end{eqnarray}
where
\begin{equation}
	\textbf{O}_{a\Phi}^{\psi} \equiv i \left(\bar{Q}_L \textbf{Y}_U \tilde{\Phi} u_R - \bar{Q}_L \textbf{Y}_D \Phi d_R - \bar{L}_L \textbf{Y}_E \Phi e_R \right) \dfrac{a}{f_a} + h.c.
\end{equation}
Here $\tilde{\Phi} = i \sigma^2 \Phi^\star$ and $\textbf{Y}_U, \textbf{Y}_D, \textbf{Y}_E$ represent 3$\times$3 
matrices in flavour space, encapsulating the Yukawa couplings for up quarks, down quarks and charged leptons, respectively.\\
It is important to note that, while our analysis focuses on a purely bosonic Lagrangian by setting the derivative couplings to fermions 
($\textbf{C}_{F}$) to zero, the effects of top quark couplings to ALP can still be addressed through the inclusion of certain bosonic operators. 
In particular, as demonstrated in Appendix A of Ref. \cite{i41}, the operator $\textbf{O}_{a\Phi}^{\psi}$, 
which couples the ALP to the SM Higgs doublet,  induces interactions that are structurally similar to the direct coupling 
of the ALP to top quarks, characterized by the coupling $c_{t}$. The study shows that the magnitude and phenomenological impact of $c_{t}$
can be effectively captured by the coupling $c_{a\Phi}$. Therefore, our focus on the bosonic sector, including $\textbf{O}_{a\Phi}^{\psi}$
still allows for a meaningful exploration of ALP interactions that might otherwise be influenced by fermion couplings, 
particularly in processes involving the top quark.

The Lagrangian described in Eq.\ref{LagALP2} has been implemented into \texttt{FeynRules} \cite{Alloul:2013bka}, 
following the conventions and notation outlined in Ref.\cite{i12}. Subsequently, the resulting Universal Feynman Output (UFO)\cite{Degrande:2011ua}
 model is integrated into \texttt{MadGraph5\_aMC@NLO} \cite{Alwall:2011uj} for the purpose of conducting numerical calculations
  for cross sections and event generation.\\
Given the potential interactions between the ALP and SM gauge bosons and fermion as outlined in Lagrangian
 Eq.\ref{LagALP1}, various decay modes are possible for ALPs depending on their mass. For ALP with mass less 
 than $2m_e$, the only permissible decay mode is the ALP decays to photons, 
 denoted as $a\rightarrow\gamma \gamma$. The decay width of the ALP to two photons at the leading order is given by \cite{Bauer:2017ris}:
\begin{equation}
	\Gamma(a\rightarrow\gamma\gamma) = \dfrac{m_a^3}{4 \pi} \left(\dfrac{c_{\gamma\gamma}}{f_a}\right)^2.
\end{equation}  
As the decay width of an ALP to two photons is proportional to $m_a^3$, for very light ALPs, the $\Gamma_a$ tends to zero. 
Consequently, the ALP will not decay inside the detector and will manifest itself as missing energy. For heavier ALPs with $m_a > 2m_e$, 
they can decay into electron pairs. In general, for $m_a > 2m_\ell$, ALPs can decay into lepton pairs. 
Moreover, ALPs are capable of decaying into partons: $a\rightarrow gg$ or $a \rightarrow q \bar{q}$. 
At the hadronic level, for ALPs with $m_a > m_\pi$, the possible decay modes are $a \rightarrow 3 \pi^0$ and $a \rightarrow \pi^+ \pi^- \pi^0$.\\
The probability of ALP decay inside the detector is given by:
\begin{equation}
	P_a^{det} = 1 - e^{L_{det}/L_a},
\end{equation}   
where $L_{det}$ is the transverse distance of the detector component from the collision point and $L_a$ is ALP decay length which is obtained to be:
\begin{equation}
	L_a = \gamma \beta \tau = \dfrac{|\vec{p}_a|}{m_a} \dfrac{1}{\Gamma_a},
\end{equation} 
where $\gamma$ represents the ALP Lorentz factor, $\tau$ is ALP proper lifetime, $\vec{p}_a$ 
denotes ALP three-momentum in each event and $\Gamma_a$ is ALP total width. 
In this study, the probability of the ALP decaying outside the detector is evaluated on 
an event-by-event basis. For each event, it is imperative that the ALP does not decay within the detector volume. 
This ensures that the ALP escapes detection, manifesting as missing energy.

In the following analyses,  to ensure the validity of the effective Lagrangian given by Eq.\ref{LagALP1}, 
it's essential that the suppression scale, $f_a$,
 in the current model is much larger than the typical energy scale, $\sqrt{\hat{s}}$, of the processes we're studying. 
 Ideally, we should have $ \sqrt{\hat{s}} < f_a$ for the events under consideration.
However, since the processes we're investigating result in an undetectable ALP in their final states, $\sqrt{\hat{s}}$ 
cannot be directly measured experimentally. In such cases, we can use the highest missing transverse energy data bin, 
$|\vec{p}_{\rm T,miss}|^{\text{max}}$, in our analysis to ensure the validity of the effective field theory. 
 In each event, we impose the requirement $|\vec{p}_{\rm T,miss}|^{\text{max}} < f_{a}$.

%%%%%%%%%%%%%%%%%%%%%%%%%%%%%%%%  
\section{ALP production in association with a top quark and $W$ boson}
\label{sectw}

\begin{figure}[t]
	\centering
	\includegraphics[width=0.75\linewidth]{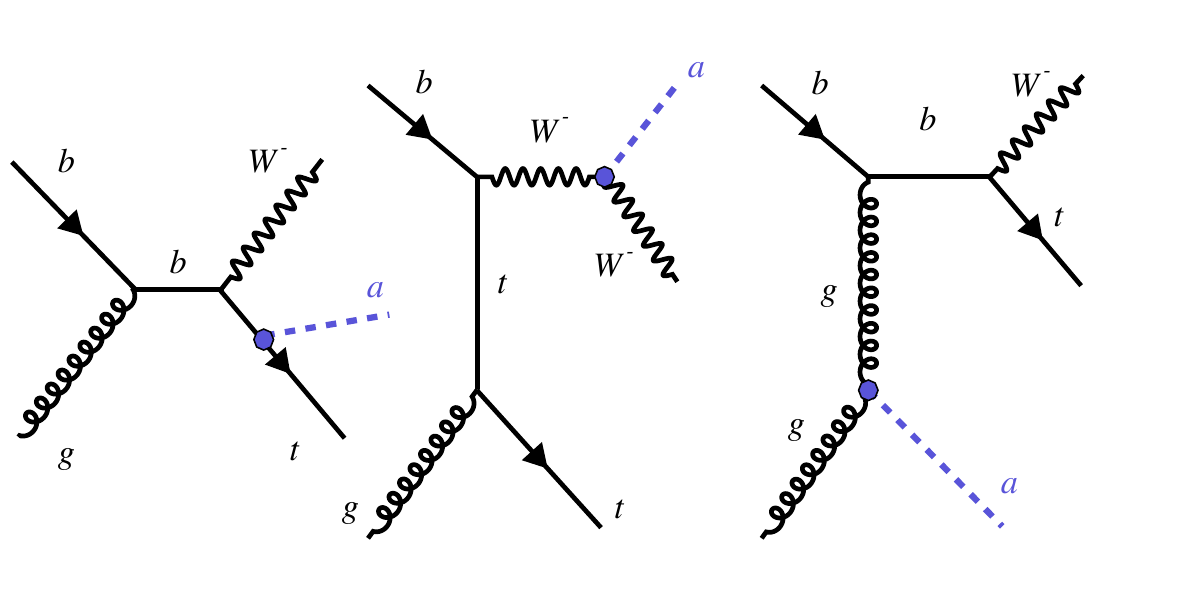}
	\caption{Representative Feynman diagrams for production of an ALP in association with a top quark and W boson.}
	\label{fig:tWax}
\end{figure}

In this section, we investigate the sensitivity of the production of an ALP in association with a top quark and a $W$ boson ($tW+$ALP). 
The Feynman diagrams representing this process are illustrated in Fig. \ref{fig:tWax}. 
As mentioned in the previous section, we consider events where the ALP does not decay within the detector, 
resulting in missing energy. Consequently, the final state is similar to the SM $tW-$channel but with a larger amount of 
missing energy expected in each event.
This process is sensitive to the couplings of ALP with gluons $c_{\tilde{G}}$,  $W$ bosons $c_{\tilde{W}}$,  and top quarks $c_{a\Phi}$.

In Ref.\cite{cmstw},  a measurement of inclusive and normalized differential $tW$ 
production cross sections at a center-of-mass energy of $ \sqrt{s} = 13$ TeV,  focusing
on the dilepton final states $(e^\pm \mu^\mp)$, are presented. The data used in this study was collected
 with the CMS detector between 2016 and 2018, amounting to an integrated luminosity of $138$ fb$^{-1}$.
After applying a baseline event selection, the number of $t\bar{t}$ background events is significantly higher 
than the $tW$ signal across all event categories. The region with the best signal-to-background 
ratio is the 1j1b region (one jet which should be a b-tagged jet), 
primarily composed of $tW$ and $t\bar{t}$ events. This region is combined 
with the 2j1b region (two jets from which one b-tagged jet), 
which also has a notable $tW$ contribution to extract the $tW$ signal.
Two independent boosted decision trees are trained, one for the 1j1b region and another for the 2j1b region, 
to distinguish between the $tW$ signal and $t\bar{t}$ background.
The measured distributions are unfolded from the detector level to the particle level. 
The differential cross section is evaluated as a function of various physical observables, including the following:
\begin{itemize}
    \item $ p_z(e^\pm, \mu^\mp, j)$: the longitudinal momentum component of the $e+\mu$ + jet system.
    \item $ \Delta \phi(e^\pm, \mu^\mp)$: the azimuthal angle difference between the electron and muon.
    \item $m_T(e^\pm, \mu^\mp, j, \vec{p}_{\text{miss}}^T)$: the transverse mass of the $e+\mu$ + jet + $\vec{p}_{\text{T,miss}}$ system defined as:
    \begin{eqnarray}
    m_{T} = \sqrt{(|\vec{p}_{\text{T},e}| + |\vec{p}_{\text{T},\mu}| + |\vec{p}_{\text{T},j}| + |\vec{p}_{\text{T,miss}}|)^{2} -
     |\vec{p}_{\text{T},e} + \vec{p}_{\text{T},\mu} + \vec{p}_{\text{T},j} + \vec{p}_{\text{T,miss}}|^2}.
    \end{eqnarray}
\end{itemize}

\begin{figure}[t]
	\centering 
	\includegraphics[width=0.45\linewidth]{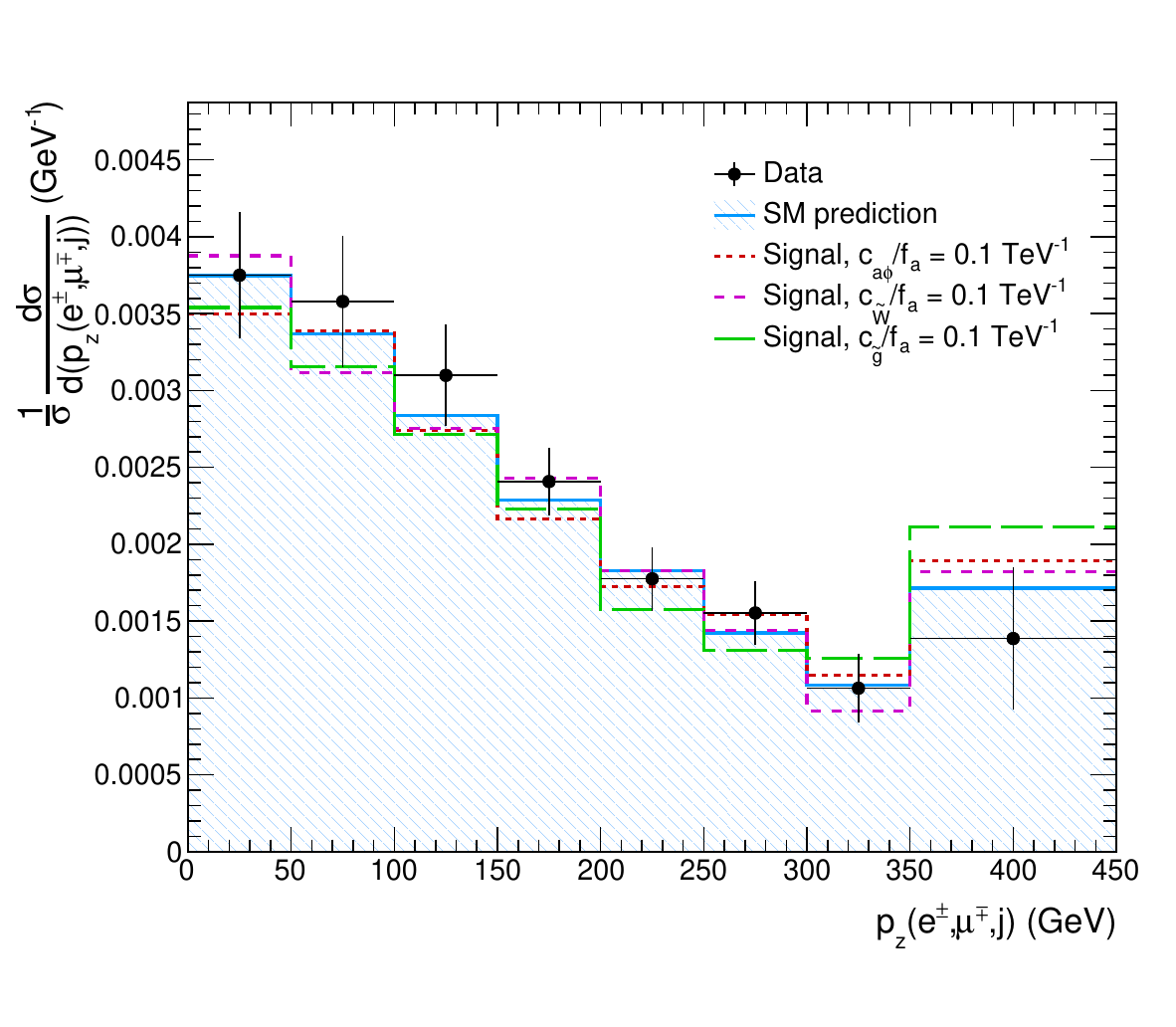}
	\includegraphics[width=0.45\linewidth]{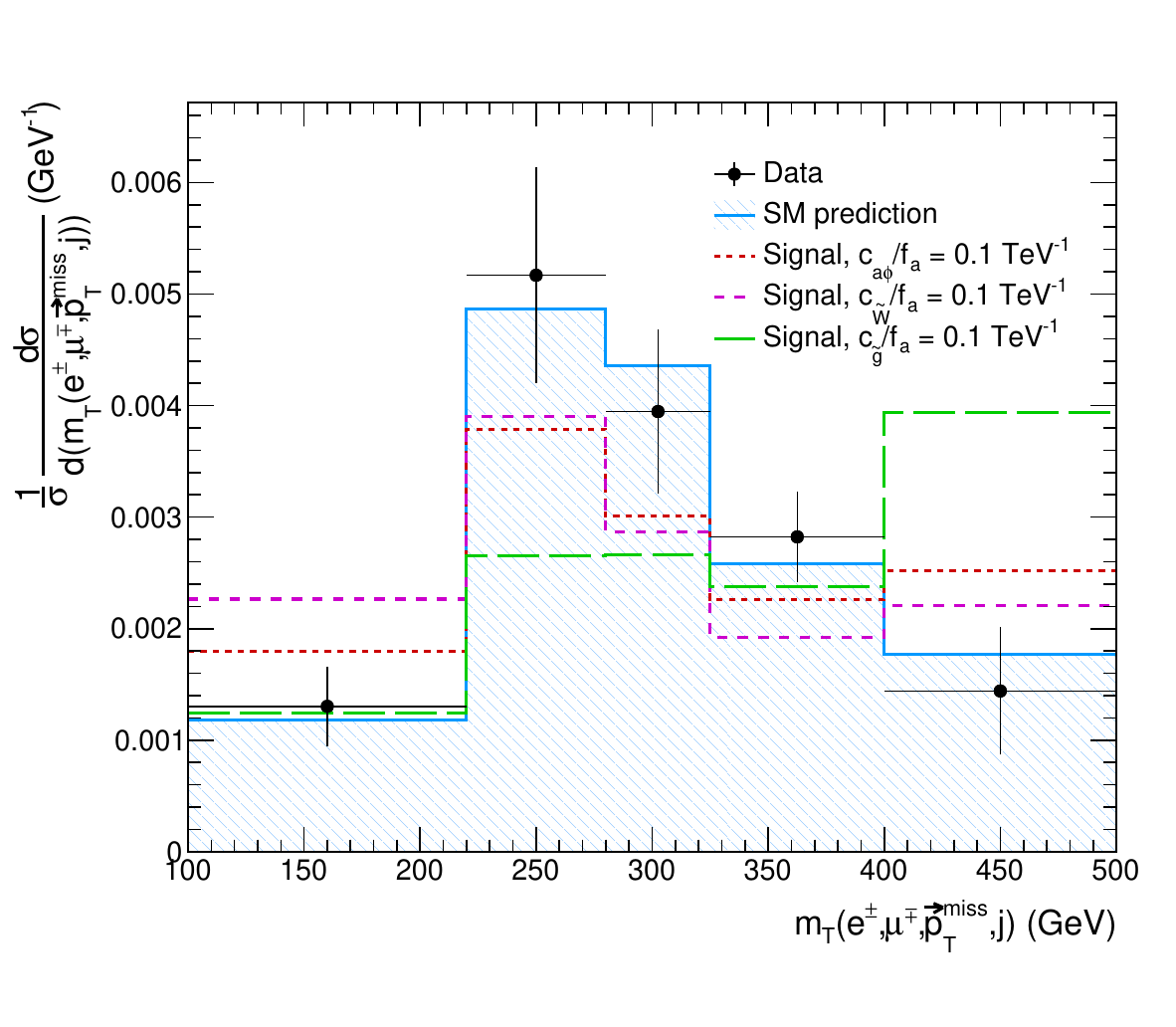}\\
	\includegraphics[width=0.45\linewidth]{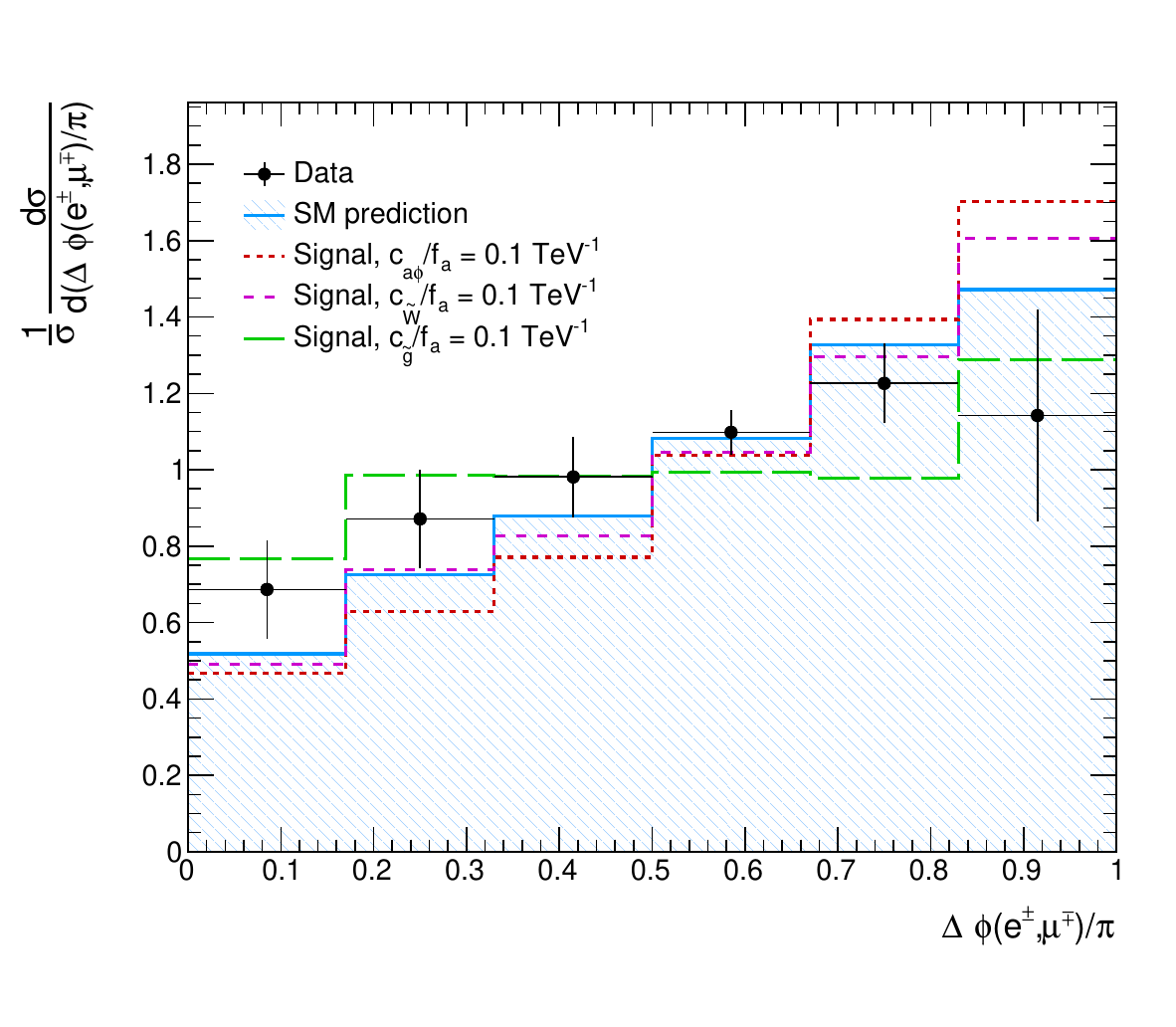}
	\caption{The normalized distribution of  the longitudinal component of the system formed by the muon, electron and jet (upper left), 
	transverse mass of the system formed by the electron, muon, 
	 jet and the missing transverse energy  (upper right),
	and difference between the angle $\phi$ between the two leading leptons (bottom)
	 are depicted for an ALP with $m_a$ = 1 MeV, SM prediction for the $tW$ process and data. The data has been taken from Ref. \cite{cmstw}.}
	\label{fig:distWax}
\end{figure}

We use the experimental measurements of these three differential distributions in the fiducial region described in
 Ref.~\cite{cmstw} as input, along with our theoretical expectations for the observables in the ALP model. 
 Subsequently, a fit is performed to extract bounds on  $c_{\tilde{G}}$, $c_{\tilde{W}}$ and $c_{a\Phi}$ couplings.
The ALP signal events are generated using \texttt{MadGraph5\_aMC@NLO}, considering 
one coefficient non-zero at a time. Subsequently, the events are passed to \texttt{Pythia 8} \cite{Sjostrand:2006za} 
for showering, hadronization and decay of the stable particles. The search involves events featuring an electron 
and a muon resulting from the decay of $W$ bosons, along with one jet and one b-jet originating from the decay of the top quark. 
Additionally, there is significant missing transverse energy corresponding to two neutrinos and ALP.\\ 
The fiducial region where the differential measurements has been made is described as follow \cite{cmstw}.
For leptons, it is required to have at least two leptons, with the first two leading leptons have opposite charge and flavor 
(one electron and one muon). We set the following criteria: the transverse momentum of the leading lepton must exceed 25 GeV,
 while for the subsequent leptons, it should be above 20 GeV, with $|\eta| < 2.4$. For electrons specifically, 
 we also exclude those with $1.4442 < |\eta| < 1.566$.
Moreover,  the minimum invariant mass between any two lepton pairs is required to be greater than 20 GeV. 
For jet selection, we need one b-jet with $p_{T,j} > 30$ GeV and $|\eta| < 2.4$. 
Jets with $\Delta R(\rm lepton,jet) < 0.4$ are also rejected, where $\Delta  R$ is defined as $\Delta R = \sqrt{\Delta \eta^2 + \Delta \phi^2}$ 
and represents the angular separation between a jet and a lepton. 
Events with any additional loose jets are rejected. 
The loose jets are defined as those with $p_{T} > 20 $ GeV and $p_{T} < 30$ GeV and $|\eta| < 2.4$. 
 After applying these requirements, the normalized distributions of the differences 
 between the angles $\phi$ of the two leading leptons, 
 $p_z(e^{\pm},\mu^{\mp},j)$, and $m_{T}(e^{\pm},\mu^{\mp},\vec{p}^{miss}_{T},j)$ are 
 depicted in Figure \ref{fig:distWax}. 
 The distributions for ALP signal events, with one coupling non-zero at a time and set to 0.1, are also presented. The ALP mass is taken as $m_a$ = 1 MeV.
 
From Fig. \ref{fig:distWax}, it is evident that 
in the $tW$ production process, and notably in its variant involving an ALP ($tW$+ALP), 
the $p_z$ distribution exhibits a pronounced peak at zero. This distinctive feature indicates that the 
$tW$+ALP system is not subject to substantial boosting along the beam axis. 
Such a characteristic implies that the majority of events result in the $tW$+ALP
system being produced with minimal longitudinal momentum, effectively being nearly at
rest in the transverse plane. Furthermore, the production predominantly occurs near the central region of the detector. 
This observation offers valuable insights into the kinematics of the $tW$+ALP system and underscores the importance of 
considering this distribution to probe the ALP model parameter space.

As depicted in the upper right plot of Fig. \ref{fig:distWax}, 
the transverse mass ($ m_T $) distribution in the single top  $tW$ + ALP in the presence of $c_{\tilde{G}}$ 
tends to reside at higher values compared to the SM scenario. 
This elevated peak suggests that the $tW$ system, influenced by the presence of the ALP, 
exhibits increased transverse momentum. Consequently, the enhanced peak in the $tW$ + ALP 
system is anticipated to yield more stringent constraints on the couplings of the ALP with the gluons.

To determine the upper limit on ALP couplings, 
We compute uncorrelated $\chi^2$ functions of the ALP couplings and mass for the differential cross section measurements 
of $p_{z}, \Delta\phi$, and $m_{T}$ defined as follows:
\begin{eqnarray}\label{chi2}
	\chi^2(c_{XX}/f_a,m_a) = \sum_{i \in bins} \dfrac{\left(\text{data}(i) -\text{bkg}(i)- \text{pred}(c_{i;XX}/f_a,m_a) \right)^{2}}{\delta_{i}^2},
\end{eqnarray} 
where data($i$) represents the measured differential distribution,  pred($i$), and bkg($i$)
 are the differential distribution of the ALP prediction and the SM background in the $i$th bin, respectively. 
 $\delta_i$ denotes the systematic and statistical uncertainties. 
 The uncertainties, the measured distribution from the data, and the SM prediction are obtained from the CMS measurement of the 
 differential cross sections for the production of single top quarks in association with a $W$ boson at the LHC with a center-of-mass energy 
 of 13 TeV and an integrated luminosity of 138 fb$^{-1}$ \cite{cmstw}.
The expected and observed upper limit on $c_{\tilde{W}}/f_a$, $c_{\tilde{G}}/f_a$ and $c_{a\Phi}/f_a$ at 95\% CL for $m_a$ = 1 MeV 
are presented in Table \ref{tab:RtWax_ex}. Table \ref{tab:RtWax_HLLHC} 
presents the constraints on $c_{\tilde{W}}/f_a$, $c_{\tilde{G}}/f_a$ and $c_{a\Phi}/f_a$ at $95\%$ CL for the HL-LHC where we considered
systematic uncertainties half of the systematic uncertainties of the current measurements by the CMS experiment.

\begin{table}[H]
	\renewcommand{\arraystretch}{1.6}
	\centering
	\resizebox{0.6\textwidth}{!}{ 
		\begin{tabular}{|c|c|c|c|}
			\hline
			   \multicolumn{4}{|c|}{ Expected (Observed) upper limits, 138 fb$^{-1}$ } \\ \hline 
			Coupling  in GeV$^{-1}$     &  $\Delta \phi(e,\mu)$ & $p_z(e^{\pm},\mu^{\mp},j)$ &  $m_{T}(e^{\pm},\mu^{\mp},\vec{p}^{miss}_{T},j)$ \\ \hline
			$|c_{\tilde{W}}/f_a|$ &  0.208 (0.307) &  0.012 (0.138) & 0.008 (0.246)    \\ \hline
			$|c_{\tilde{G}}/f_a|$ &  0.074 (0.110) &  0.004 (0.046)   &  0.003 (0.097)  \\ \hline
			$|c_{a\Phi}/f_a|$   &    0.407 (0.605) &  0.023 (0.267)  &  0.016 (0.500)   \\ \hline 
	\end{tabular}}
    \caption{The $95\%$ CL expected and observed upper limits on $c_{\tilde{W}}/f_a$, $c_{\tilde{G}}/f_a$ and $c_{a\Phi}/f_a$ 
    for an ALP with mass $m_a$ = 1 MeV from fits to $p_{z}, \Delta\phi$ and $m_{T}$ distributions. The fit on $p_{z}$ has been performed on the region
    with $p_{z} > 150$ GeV.}
    \label{tab:RtWax_ex}
\end{table}

\begin{table}[H]
	\renewcommand{\arraystretch}{1.6}
	\centering
	\resizebox{0.6\textwidth}{!}{ 
		\begin{tabular}{|c|c|c|c|}
			\hline
			   Coupling in GeV$^{-1}$ & $\Delta \phi(e,\mu)$ & $p_z(e^{\pm},\mu^{\mp},j)$ &  $m_{T}(e^{\pm},\mu^{\mp},\vec{p}^{miss}_{T},j)$ \\ \hline
			$|c_{\tilde{W}}/f_a|$ &  0.132 &     0.007 &  0.005 \\ \hline
			$|c_{\tilde{G}}/f_a|$  & 0.047 &  0.002   &  0.002 \\ \hline
			$|c_{a\Phi}/f_a|$    &  0.259 &    0.014  & 0.011  \\ \hline
	\end{tabular}}
	\caption{The $95\%$ CL expected upper limits on $c_{\tilde{W}}/f_a$, $c_{\tilde{G}}/f_a$ and $c_{a\Phi}/f_a$ for 
	an ALP with mass $m_a$ = 1 MeV for the HL-LHC with an integrated luminosity of  3 ab$^{-1}$.}
	\label{tab:RtWax_HLLHC}
\end{table}

In probing the ALP couplings with gluons, $W$ boson, and top quarks through the $tW+$ALP channel, 
our analysis revealed intriguing insights into the sensitivity of various kinematic variables to these couplings.
At the HL-LHC, among these observables, $\Delta \phi(e,\mu)$ provides the weakest constraints across all couplings, 
with upper limits of 0.132 GeV$^{-1}$ for $|c_{\tilde{W}}/f_a|$, 0.047 GeV$^{-1}$ for $|c_{\tilde{G}}/f_a|$, and 0.259 GeV$^{-1}$ for $|c_{a\Phi}/f_a|$. 
In comparison, the $p_z(e^{\pm},\mu^{\mp},j)$ observable yields significantly tighter constraints, with upper limits of 0.007 GeV$^{-1}$ for 
$|c_{\tilde{W}}/f_a|$, 0.002 GeV$^{-1}$ for $|c_{\tilde{G}}/f_a|$, and 0.014 GeV$^{-1}$ for $|c_{a\Phi}/f_a|$. 
Similarly, the $m_{T}(e^{\pm},\mu^{\mp},\vec{p}^{miss}_{T},j)$ observable also shows strong constraints, slightly 
less stringent than $p_z(e^{\pm},\mu^{\mp},j)$ but still considerably better than $\Delta \phi(e,\mu)$, 
with upper limits of 0.005 GeV$^{-1}$ for $|c_{\tilde{W}}/f_a|$, 0.002 GeV$^{-1}$ for $|c_{\tilde{G}}/f_a|$, and 0.011 GeV$^{-1}$ for $|c_{a\Phi}/f_a|$. 
Overall, while $\Delta \phi(e,\mu)$ is the least constraining observable for these couplings, $p_z(e^{\pm},\mu^{\mp},j)$ 
and $m_{T}(e^{\pm},\mu^{\mp},\vec{p}^{miss}_{T},j)$ provide much tighter limits, with $p_z(e^{\pm},\mu^{\mp},j)$ 
being marginally the most restrictive across all couplings.

The observed constraints presented in Table \ref{tab:RtWax_ex} are looser than the expected ones, mainly because of the large uncertainties 
in the CMS experiment measurements, as shown by the error bars on the data in the distributions. 
For example, in the $p_z$ distribution, even though the shape of the SM prediction 
is quite different from the ALP model, the data uncertainty bars encompass the ALP expectation. 
This leads to less stringent observed limits with respect to the expected ones.

%%%%%%%%%%%%%%%%%%%%%%%%%%%%%%%%  
\section{Probing the ALP parameter space through the spin correlations and $\Delta\phi$ distribution in $t\bar{t}$ events}
\label{secspin}

The spin information of top quarks, being unstable, is inferred through their decay products. 
However, not all decay products equally carry the spin information. 
Charged leptons from the leptonic decays of $W$ bosons, 
in particular, capture almost the entirety of the spin information of 
their parent top quark \cite{s1,s2,s3}. This characteristic, 
combined with the ease of identification and reconstruction of charged
 leptons in collider experiments, makes them pivotal for studying spin correlations in   $t\bar{t}$ events.

In events where both $W$ bosons decay leptonically, commonly known as the 
dilepton channel, the angular distributions of the charged leptons serve as valuable 
observables for spin correlation studies. The most straightforward observable is the 
absolute azimuthal opening angle between the two charged leptons \cite{s4}, 
measured in the laboratory frame in the plane transverse to the beam line. This angle is denoted by $ \Delta\phi_{\ell\ell}$.
Following this an asymmetry is defined as a single observable to quantify 
the amount of spin correlation as follows:

\begin{equation}
	A_{|\Delta\phi_{\ell\ell}|} = \dfrac{N(|\Delta\phi_{\ell\ell}| > \pi/2) - N(|\Delta\phi_{\ell\ell}| < \pi/2)}{N(|\Delta\phi_{\ell\ell}| > \pi/2) + N(|\Delta\phi_{\ell\ell}| < \pi/2)},
\end{equation}
The SM prediction for $A_{|\Delta\phi_{\ell\ell}|} $ is $0.108^{+0.009}_{-0.012}$
and the CMS experiment measurement at the center-of-mass energy of 13 TeV
with an integrated luminosity of 35.9 fb$^{-1}$ is $0.103 \pm 0.008$ \cite{CMS:2019nrx}.

\begin{figure}[ht]
	\centering
	\includegraphics[width=0.75\linewidth]{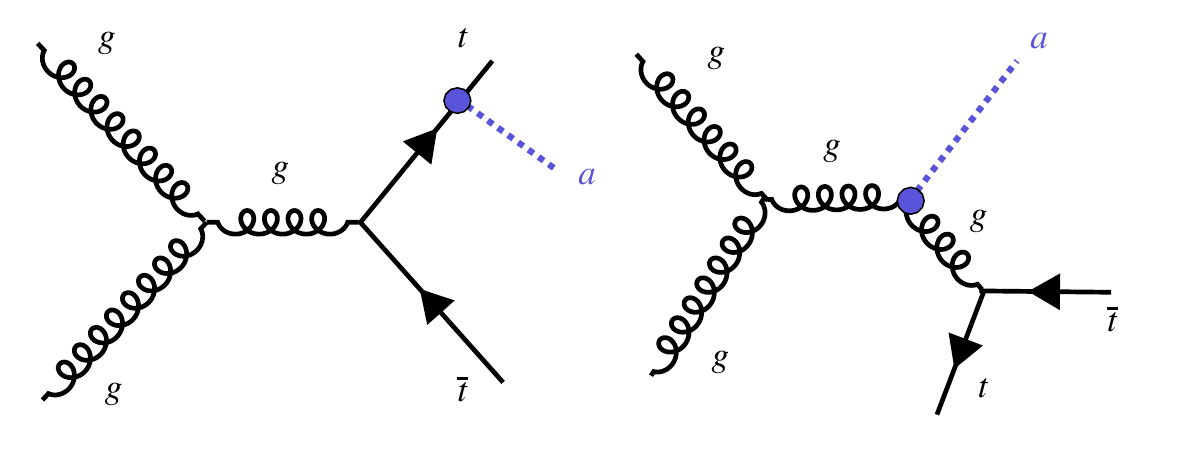}
	\caption{Representative Feynman diagrams for production of an ALP in association with top quark pairs at the LHC.}
	\label{fig:ttaxxx}
\end{figure}

When the ALP is stable and escapes the detector without decaying, 
the $t\bar{t}$+ALP events closely resemble those of standard $t\bar{t}$ production. We make use of the
 similarity in signatures and perform a $\chi^{2}$ fit on the measured $ \Delta\phi_{\ell\ell}$ distribution by the
CMS experiment to extract limits on the ALP couplings with gluons and top quarks.
The Feynman diagrams illustrating this process are presented in Fig.\ref{fig:ttaxxx}. 
At production level, this measurement enables us to explore the $c_{\tilde{G}}$ and $c_{a\Phi}$ couplings.

\begin{figure}[ht]
	\centering
	\includegraphics[width=0.5\linewidth]{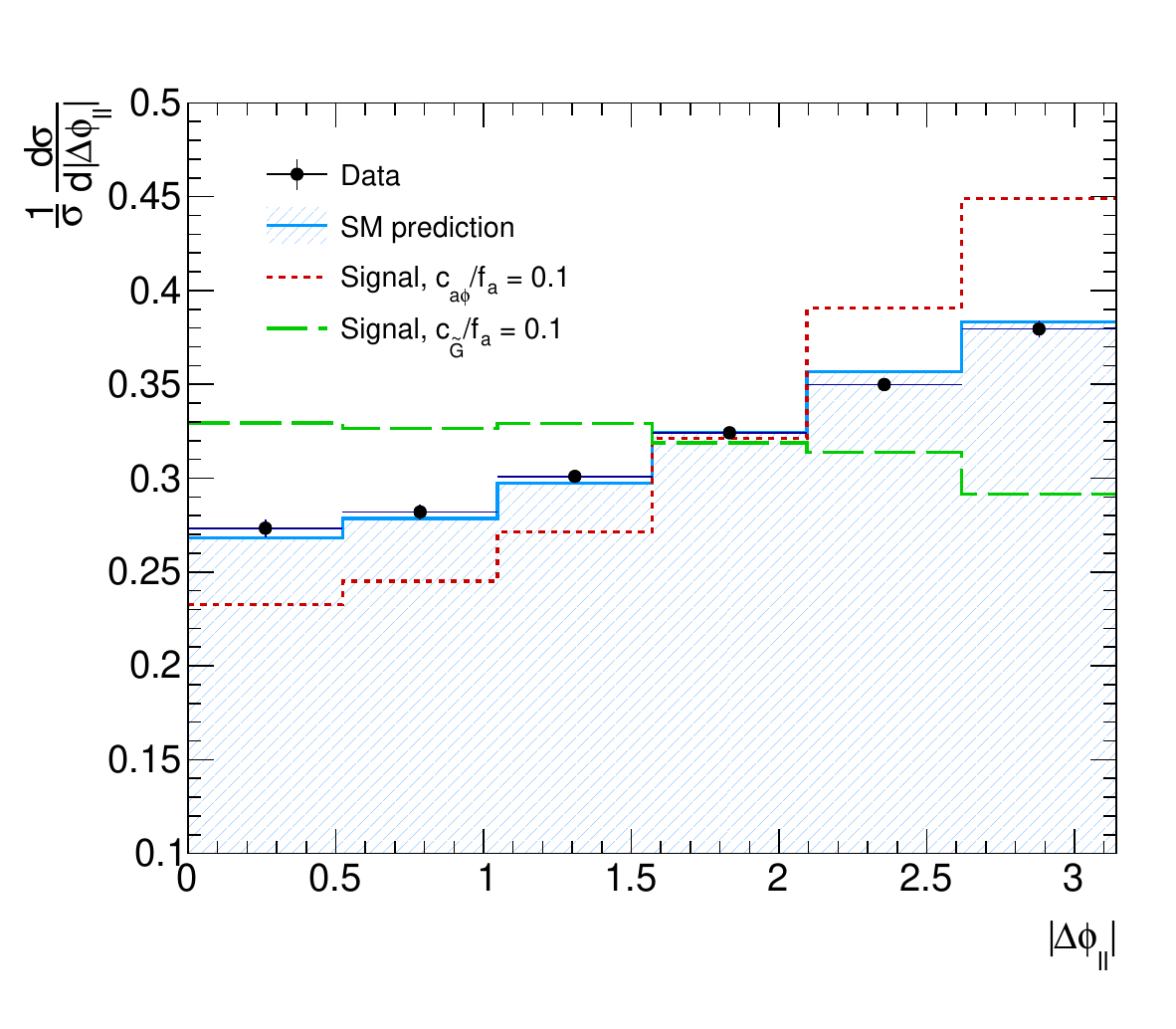}
	\caption{The distribution of the absolute azimuthal opening angle between the two charged leptons ($\Delta\phi_{\ell\ell}$) from the top quarks decay
	 for SM, data, and the ALP prediction over the SM once with $c_{\tilde{G}}/f_a = 0.1$ TeV$^{-1}$ and once with $c_{a\Phi}/f_a = 0.1$ TeV$^{-1}$. The data 
	 is taken from the CMS experiment measurement at the center-of-mass energy of 13 TeV with an integrated luminosity of 35.9 fb$^{-1}$  \cite{CMS:2019nrx}.}
	\label{fig:dphi_ttax}
\end{figure}

When an ALP is radiated off from one of the top quarks due to CP-odd interactions, 
it introduces additional complexity to the spin correlation and  $\Delta\phi_{\ell\ell}$ distributions. 
The presence of the ALP can modify the momenta and energies of the decay products, 
altering the angular distributions and correlations between the charged leptons.
In particular, the emission of an ALP  leads to a change in the momentum of the top quark, 
which in turn influences the momentum of the $W$ boson and its subsequent decay products.
As a result, the azimuthal opening angle  between the charged leptons $\Delta\phi_{\ell\ell}$  
deviates from its expected distribution in the absence of ALP radiation. 
This deviation serves as an indirect signature of the ALP interaction and is used in this section
to explore the parameter space of the ALP. 
Figure \ref{fig:dphi_ttax} displays the distribution of $\Delta\phi_{\ell\ell}$
across various scenarios: SM, observed data, and ALP predictions considering two distinct ALP scenarios. 
The ALP scenarios are represented by the coupling parameters $c_{\tilde{G}}/f_a = 0.1$ TeV$^{-1}$ and $ c_{a\Phi}/f_a = 0.1$ TeV$^{-1}$. 
The data is sourced from the CMS experiment, specifically from measurements taken at a 
center-of-mass energy of 13 TeV, with an integrated luminosity of 35.9 fb$^{-1}$  \cite{CMS:2019nrx}.

The $t\bar{t}+\text{ALP}$ signal events are generated using the methods detailed in the previous section. 
We consider the ALPs which do not decay inside the detector, manifesting itself as missing transverse momentum. 
Considering the leptonic decay channel of the top and anti-top quark, the final state comprises two oppositely 
charged leptons, two b-jets, and significant missing transverse momentum from the two neutrinos and the ALP.

We employ a $\chi^{2}$ test similar to Eq.\ref{chi2} on the $\Delta\phi_{\ell\ell}$
distribution to extract constraints on the couplings of ALPs with
 top quarks ($c_{a\Phi}/f_a$) and gluons ($c_{\tilde{G}}/f_a$).  
The observed upper limit on $c_{\tilde{G}}/f_a$ and $c_{a\Phi}/f_a$ at 95\% CL for $m_a = 1$ MeV and $\mathcal{L} = 35.9 $ fb$^{-1}$ are found to be:
\begin{eqnarray}
	\dfrac{c_{\tilde{G}}}{f_a} \leq 0.012 \: \text{GeV}^{-1} \quad , \quad \dfrac{c_{a\Phi}}{f_a} \leq 0.124 \: \text{GeV}^{-1}.
\end{eqnarray} 
and the expected upper limits are:
\begin{equation}
	\dfrac{c_{\tilde{G}}}{f_a} \leq 0.007 \: \text{GeV}^{-1} \quad , \quad \dfrac{c_{a\Phi}}{f_a} \leq 0.064 \: \text{GeV}^{-1},
\end{equation} 
For the HL-LHC with an integrated luminosity of $\mathcal{L} = 3000 $ fb$^{-1}$ and
 assuming a reduction of half in the statistical and systematic uncertainties, the resulting expected limits are obtained to be: 
\begin{equation}
	\dfrac{c_{\tilde{G}}}{f_a} \leq 0.004 \: \text{GeV}^{-1} \quad , \quad \dfrac{c_{a\Phi}}{f_a} \leq 0.041 \: \text{GeV}^{-1}.
\end{equation}

\begin{figure}[ht]
	\centering
	\includegraphics[width=0.9\linewidth,height=7cm]{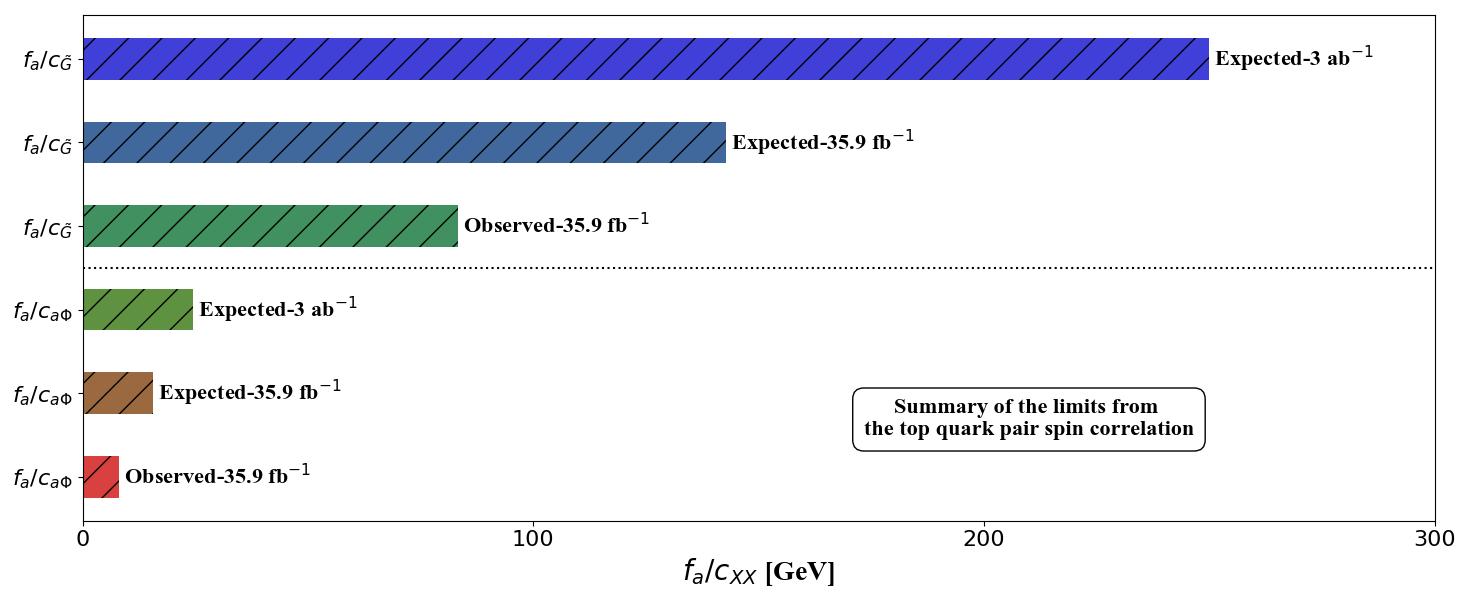} % Set width and height separately
	\caption{Summary of the constraints on the ALP couplings with top quarks and gluons derived
	 from the fit to the $\Delta\phi$ distribution in the dileptonic channel of $t\bar{t}$.}
	\label{fig:ttax}
\end{figure}

Figure \ref{fig:ttax} provides a comprehensive summary of the
constraints on the axionlike particle couplings to top quarks
and gluons, derived from a fit to the $\Delta\phi$ distribution in the
dileptonic channel of $t\bar{t}$ events. The results are depicted for
the parameters $f/c_{a\Phi}$ and $f_{a}/c_{\tilde{G}}$ considering an
integrated luminosity of 35.9 fb$^{-1}$ and a projected integrated
luminosity of 3 ab$^{-1}$.

\section{Comparison of Constraints on ALP Couplings from Different Processes}
\label{comp}

In this section, we present a comparison of the constraints on the coupling of ALPs with the top quark ($c_{a\Phi}/f_a$) and with gluons ($c_{\tilde{G}}/f_a$) 
obtained from our analysis of single top $tW$-channel production and $t\bar{t}$ spin correlation with those derived from 
loop-induced processes such as $ZZ$, $Z\gamma$, $\gamma\gamma$, di-Higgs production, and B and K meson decays.

Loop-induced processes such as $ZZ$, $Z\gamma$, $\gamma\gamma$, and di-Higgs production provide indirect 
constraints on ALP couplings via effective operators. These processes are sensitive to various ALP interactions, 
including those with gauge bosons and the Higgs boson. The  measurements of diboson and 
Higgs pair production cross-sections and decay rates at the LHC have been employed to derive constraints on ALP couplings.
These processes are sensitive to ALP couplings through loop diagrams involving top quarks. 
The analysis of $ZZ$ and  $Z\gamma$ productions at the LHC have constrained $c_{a\Phi}/f_a$ to values of $<  0.28$ and $< 0.088$ GeV$^{-1}$ at $95\%$ CL, respectively \cite{i41}. 
Photon pair production is another crucial probe, sensitive to ALP couplings with photons mediated via top quark loops. 
The LHC measurements of $\gamma\gamma$ production set constraints on $c_{a\Phi}/f_a$  to be $< 0.044$ GeV$^{-1}$ at $95\%$ CL \cite{i41}. 
The better sensitivity arises from the clean experimental signatures and precise theoretical predictions for these processes.
The experimental precision in this channel benefits from the excellent photon reconstruction capabilities of the LHC detectors.
The production of Higgs boson pairs from the ALP-mediate provides a unique probe of ALP interactions with the Higgs sector
which can be translated into bounds on the coupling of ALP to top quarks. Constraints from 
di-Higgs production are more stringent compared to diboson processes. 
Current limits from LHC measurements constrain $c_{a\Phi}/f_a$  to be $< 0.0046$ GeV$^{-1}$ at $95\%$ CL \cite{Esser:2024pnc}.
The latest measurement of   $K \to \pi\nu\bar{\nu}$ from NA62 and the current best limit for B-decays from BaBar
 lead to constraints on $c_{a\Phi}/f_a$ of $< 0.00028$ and $< 0.00000147$ GeV$^{-1}$ at $95\%$ CL, respectively \cite{i41}.

Direct constraints on the ALP-top coupling have been obtained from the ALP production 
in association with a $t\bar{t}$ pair. 
A reinterpretation of the Run II ATLAS search for top squarks in events with two leptons, b-jets, and missing transverse energy (MET) 
at $\sqrt{s} = 13$ TeV and $L = 139$ fb$^{-1}$, provides $|c_{a\Phi}/f_a| < 0.0018$ \cite{i41}.

The constraints on ALP couplings with gluons, $c_{\tilde{G}}/f_a$, have been derived from various decay processes. 
The measurement of $K^+ \to \pi^+\nu_\ell\bar{\nu}_\ell$ yields a limit of $|c_{\tilde{G}}/f_a| < 0.00000769$ GeV$^{-1}$. 
For $B \to K^*\nu_\ell\bar{\nu}_\ell$, the constraint is $|c_{\tilde{G}}/f_a| < 0.0021$ GeV$^{-1}$. The decay process 
$\pi^+ \to ae^+\nu_e$ provides a limit of $|c_{\tilde{G}}/f_a| < 0.00454$ GeV$^{-1}$. 
Among the indirect probes, the loosest constraint comes from $B_s - \bar{B}_s$ mixing, which sets a limit of $|c_{\tilde{G}}/f_a| < 2.352$ GeV$^{-1}$.

Our results provide competitive constraints on ALP couplings, particularly when considering direct production processes. 
The constraints on $c_{\tilde{G}}/f_a$ and $c_{a\Phi}/f_a$ from the single top $tW$-channel and $t\bar{t}$ spin correlation 
analyses are robust and offer complementary information to the constraints derived from loop-induced processes. 
While loop-induced processes provide stringent indirect constraints, they rely on effective operators and are sensitive 
to ALP interactions with multiple particles, which can complicate the interpretation.
In contrast, our direct probes focus specifically on the ALP-top and ALP-gluon couplings, yielding clear and interpretable results. 
The observed and expected limits we obtained for $c_{\tilde{G}}/f_a$ and $c_{a\Phi}/f_a$ are competitive with and, in some cases, 
more stringent than those from indirect methods. Additionally, our projections for the HL-LHC show significant improvement, 
emphasizing the potential of direct searches in future collider experiments.

Overall, our results, derived from single top $tW$-channel and $t\bar{t}$ spin correlation analyses, 
highlight the importance of combining direct and indirect probes to achieve a comprehensive understanding of ALP couplings. 
The synergy between different analysis channels enhances the sensitivity to new physics, ensuring robust and reliable constraints on the top-ALP coupling. 
Future improvements in experimental precision and theoretical calculations will further refine these constraints, offering deeper insights into the nature of ALPs 
and their interactions with the SM particles.

\section{Concurrent Limits of ALP Interactions with Gluons and Top Quark}
\label{2Dlimit}

In this section, we investigate the two ALP couplings $c_{\tilde{G}}/f_a$ and $c_{a\Phi}/f_a$
concurrently using the total cross sections of the $tW$ and $t\bar{t}$ processes. 
The measurements of these processes are interpreted as $tW$+ALP and $t\bar{t}$+ALP, respectively, 
with the requirement that the ALP does not decay inside the detector, thus appearing as missing momentum.

Simultaneously probing the couplings $c_{\tilde{G}}/f_a$ and $c_{a\Phi}/f_a$
offers several advantages over examining each coupling individually. 
Firstly, this approach provides a more comprehensive exploration of the ALP parameter space, 
allowing a wider range of potential ALP interactions to be investigated within a single analysis. 
This is useful given the variety of theoretical models which can predict ALPs with different coupling 
patterns to SM particles.  Secondly, probing multiple couplings at the same time facilitates the
 identification of potential correlations or interdependencies between different ALP interactions, 
offering valuable insights into the underlying theoretical framework.

\begin{figure}[h]
	\centering
	\includegraphics[width=0.5\linewidth]{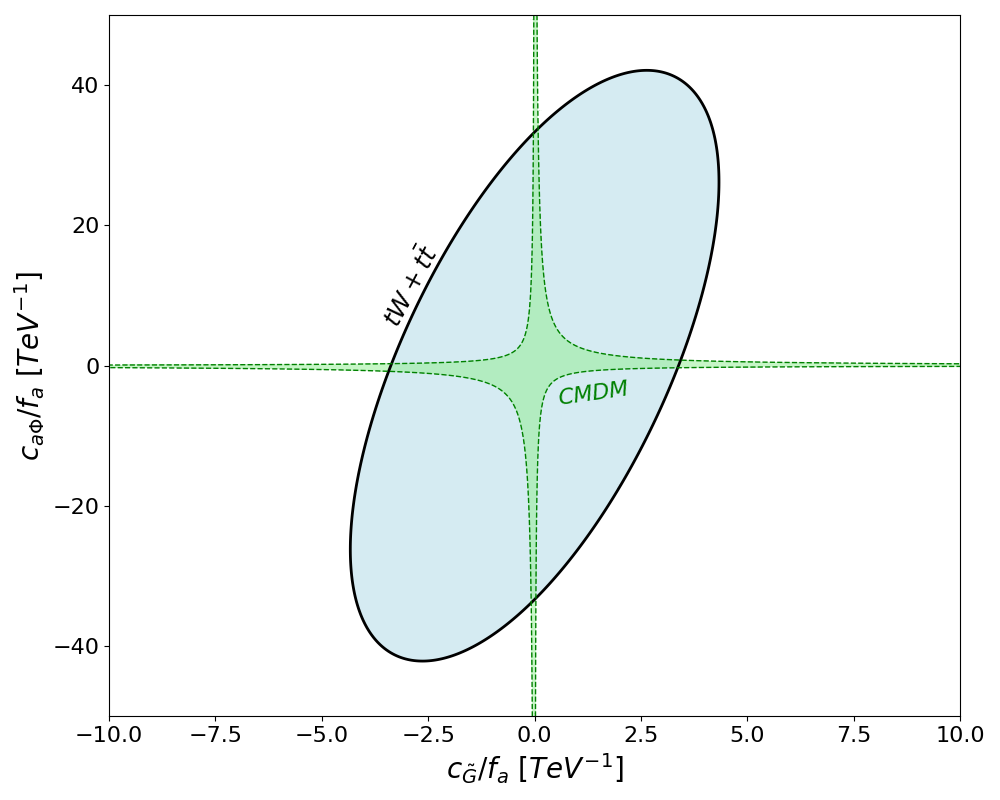}
	\caption{The $95\%$ CL allowed region in the ($c_{\tilde{G}}/f_a$,$c_{a\Phi}/f_a$) plane  using the combination of the total cross sections of $tW$ and
	$t\bar{t}$ channels. The limits from the CMDM is presented for comparison in green as dashed. The bounds have been derived for $m_{a} = 1$ MeV.}
	\label{fig:2d}
\end{figure}

To derive constraints on these couplings, we make use the measured cross sections of the $tW$ and $t\bar{t}$
processes from Ref.~\cite{cmstw} and Ref.~\cite{atlastt}, respectively. A simultaneous $\chi^{2}$ 
analysis of the inclusive cross sections of the $tW$ and $t\bar{t}$ processes is performed to 
extract the two-dimensional contour on $(c_{\tilde{G}}/f_a, c_{a\Phi}/f_a)$. 
The allowed region at $95\%$ CL is presented in Figure \ref{fig:2d} as a black ellipse. 

As depicted in the figure, the sensitivity is greater for $c_{\tilde{G}}/f_a$. This  sensitivity is 
mainly due to the dominant contribution from the gluon-gluon fusion process in $t\bar{t}$+ALP production, 
which is enhanced by the large gluon parton distribution function in the proton.

In addition to the concurrent limits derived from the total production rates, 
we present a comparison with the simultaneous bounds obtained from the
 chromo-magnetic dipole moment (CMDM) of the top quark, as discussed in
  our previous work \cite{i40}. 
  As demonstrated in Ref.~\cite{i40}, ALPs contribute to the top quark CMDM according to the following expression:

\begin{eqnarray}
\tilde{a}_{t} = \frac{c_{\tilde{G}}\times c_{a\Phi} \times m_{t}^{2}}{\pi^{2} \times f_{a}^{2}} \times (\frac{3}{4} + \log(\frac{\Lambda}{m_{a}})). 
\end{eqnarray}

The parameter \(\Lambda\) in the numerator of the logarithms, which arises from loop divergences, 
can be naturally assumed to be equal to the new physics scale. In the ALP model, 
$\Lambda$ can be assumed  to be equal to $f_a$. 
Utilizing the upper limits on the top quark CMDM allows for the constraint of the ALP model parameters. 
Based on the CMS experiment limits at 95\% CL on the CMDM of the top quark, $-0.043 \leq \tilde{a}_t \leq 0.117$  \cite{cmdm}, 
the constraints on $(c_{\tilde{G}}/f_a, c_{a\Phi}/f_a)$ are derived. 
The two-dimensional contour from the CMDM analysis is displayed as the dashed green curve in Figure \ref{fig:2d}. 
As observed in Figure \ref{fig:2d}, the indirect limits from CMDM and the limits from the inclusive 
cross sections are complementary, with each providing sensitivity to different regions of the parameter space.

%%%%%%%%%%%%%%%%%%%%%%%%%%%%%%%%  	

%%%%%%%%%%%%%%%%%%%%%%%%%%%%%%%%   
\section{Conclusions}
\label{secsummary}

Exploring the phenomenology of light axion-like particles (ALPs) is crucial for beyond the SM investigations, 
as exemplified by studies at the LHC. The LHC provides a rich environment for processes sensitive to ALP interactions, 
particularly with electroweak bosons, gluons, and the top quark. 
In this study, we probe the ALP parameter space by considering the pure bosonic effective Lagrangian up to dimension five, 
describing ALP interactions with SM fields. 
We present collider limits on ALP model parameters from the $tW+\text{ALP}$
 channel using CMS experiment measurements and project prospects at the HL-LHC.
 Differential measurements, such as the longitudinal momentum component of the  $e^\pm \mu^\mp j$ system ($p_z(e^\pm, \mu^\mp, j)$), 
 the azimuthal angle difference between the electron and muon ($ \Delta \phi(e^\pm, \mu^\mp) $), 
 and the transverse mass of the $ e^\pm \mu^\mp j+\vec{p}_{\text{T,miss}}$ system 
 ($ m_T(e^\pm, \mu^\mp, j, \vec{p}_{\text{miss}}^T)$) in the $tW+\text{ALP}$ channel, 
 emerge as promising differential observables to probe ALP couplings with $W$ bosons,
 gluons, and top quarks. 

We also investigate the $t\bar{t}$+ALP channel where both $W$ bosons from top quarks decay leptonically.
The angular distributions of the charged leptons in this channel prove to be a valuable observable 
for studying ALP couplings. 
A fit to the measured azimuthal opening angle between 
the charged leptons is performed, providing constraints on the ALP couplings with gluons and top quarks.
The measured constraints are found to be $|c_{a\Phi}/f_a| \leq 0.064 \, \text{GeV}^{-1} $
and $|c_{\tilde{G}}/f_a| \leq 0.007 \, \text{GeV}^{-1} $  for $m_a = 1$ MeV.

Overall, our analysis demonstrates that direct constraints from single top $tW$-channel and $t\bar{t}$ spin correlation 
not only complement but also enhance the understanding of ALP couplings in the top sector. This makes our results a 
valuable contribution to the ongoing efforts in constraining ALP parameters.

%\section*{Acknowledgement}
%%%%%%%%%%%%%%%%%%%%%%%%%%%%%%%%

%%%%%%%%%%%%%%%%%%%%%%%%%%%%%%%%

\end{document}